\documentclass[prb,superscriptaddress,showpacs,twocolumn]{revtex4-1}
\usepackage[english]{babel}
\usepackage[unicode]{hyperref}
\usepackage{amsmath}
\usepackage{amsfonts}
\usepackage{amssymb}
\usepackage{graphicx}
\usepackage{epstopdf}
\usepackage[usenames,dvipsnames]{color}
\usepackage{bm}
\usepackage{caption}
\usepackage{subcaption}
\usepackage{gensymb}
\usepackage{rotating}
\usepackage{array}
\usepackage{color}
\usepackage{float}

\usepackage[margincaption,outercaption,ragged,wide]{sidecap}
\sidecaptionvpos{figure}{l}

\makeatletter
\newcommand*{\citenst}[2][]{%
  \begingroup
  \let\NAT@mbox=\mbox
  \let\@cite\NAT@citenum
  \let\NAT@space\NAT@spacechar
  \let\NAT@super@kern\relax
  \renewcommand\NAT@open{[}%
  \renewcommand\NAT@close{]}%
  \cite[#1]{#2}%
  \endgroup
}
\makeatother

\begin{document}
\title{Little-Parks Oscillations in a Single Ring in the vicinity of the Superconductor-Insulator Transition}
\author{Doron Gurovich}
\email{dmpgur@weizmann.ac.il}
\affiliation{Department of Condensed Matter Physics, The Weizmann Institute of Science, Rehovot 76100, Israel}
\author{Konstantin S. Tikhonov}
\email{tikhonov@physics.tamu.edu}
\affiliation{Department of Condensed Matter Physics, The Weizmann Institute of Science, Rehovot 76100, Israel}
\affiliation{L. D. Landau Institute for Theoretical Physics, 117940 Moscow, Russia}
\affiliation{Moscow Institute of Physics and Technology, 141700 Moscow, Russia}
\author{Diana Mahalu}
\affiliation{Department of Condensed Matter Physics, The Weizmann Institute of Science, Rehovot 76100, Israel}
\author{Dan Shahar}
\affiliation{Department of Condensed Matter Physics, The Weizmann Institute of Science, Rehovot 76100, Israel}
\date{\today}

\begin{abstract}
We present results of measurements obtained from a mesoscopic ring of a highly disordered superconductor. Superimposed on a smooth magnetoresistance
background we find periodic oscillations with a period that is independent of the strength of the magnetic field. The period of the oscillations is 
consistent with charge transport by Cooper pairs. The oscillations persist unabated 
for more than 90 periods, through the transition to the insulating phase, up to our highest field of 12 \nolinebreak T.
\end{abstract}
\maketitle

\section{Introduction} Transport properties of amorphous superconducting films are strongly influenced by Cooper pairing, 
Coulomb repulsion and disorder. The interplay of these effects leads to the very interesting physics of Superconductor-Insulator 
transition (SIT), which is now routinely observed in dirty metallic films\cite{gantmakher2010superconductor, goldman2008superconductor, dobrosavljevic2012conductor}. 
This quantum phase transition can be driven by variation of disorder\cite{Fiory1984}, thickness\cite{Haviland89}, magnetic field ($B$)\cite{Hebard90}, 
composition and carrier concentration\cite{Leng11}.

One of the central questions regarding the physics of the SIT is to which extent Cooper pairing is relevant in the
insulating phase terminating superconductivity. From the theoretical side, there are two complementary approaches. 
The so-called Fermionic theory of suppression of superconductivity\cite{finkel94}, being quite successful in describing the 
reduction of the transition temperature $T_c$ via Coulomb interaction, including full suppression of superconductivity, 
does not take into account the  effects of Cooper pairing in the normal, or insulating, state of the film. The alternative approach
considers competition of Anderson localization and superconductivity\cite{ma1985,kapitulnik85,bulaevskii84,bouadim2011single} and, 
contrary, admits activated transport by Cooper pairs in the insulating regime\cite{feigel10}. 

Experimentally, the importance of Cooper pairing in the insulating state can be probed by both tunnelling spectroscopy and 
transport measurements. In the first approach, one directly measures the superconducting gap in the insulating phase, which 
indicates the presence of localized Cooper pairs\cite{sacepe11,Sherman2012, Sherman2014}. 
The second approach (which we adopt in this paper) is based on specifically addressing effects, which are related to the 
crucial property of the Cooper pairs - their ability to maintain coherence at macroscopic distance. 
This idea can be traced back to one of the first indications to the importance of the Cooper-paring principle - Little-Parks 
experiment\cite{little1962observation}. Since then, a series of experiments 
was performed following the same logic\cite{gammel90,carillo10,sochnikov10,hollen11}. 

Recently, following the experiment of J. M. Valles Jr. group \cite{stewart2009enhanced}, we applied this idea 
to amorphous indium-oxide ($a$:InO) films \cite{kopnov2012little}. 
We used a self-arranged array of holes to create a sample comprised of a network of rings of a disordered superconductor. Our measurements
 demonstrated the existence of oscillations with a period consistent with elementary charge of $2e$ (Cooper pairs) in the insulating regime. 
 However, as in other experiments \cite{gammel90,carillo10,sochnikov10,hollen11},
we were able to detect only a few oscillations, due to their decay with $B$. The reason of this decay was not clear and can, 
in principle, be twofold: 
1) intrinsic effect of magnetic field, which quickly destroys spatial coherence of the Cooper pairs 
on the scale of the elementary cell of the array and 2) effect of 
fluctuating size of the individual loops of the array, which smears out oscillations at larger fields. In addition, 
it was not possible to exclude the possibility of Josephson array physics \cite{van1988experiments}.

The aim of the present work is to extend our earlier study\cite{kopnov2012little} to the case of a single ring, in order to clarify both questions. 
We concentrated on the direct vicinity of the disorder-induced SIT transition in $a$:InO. We found that oscillations not only 
exist in a single ring both below and above SIT, but persist up to the highest fields available \nolinebreak (12 T).

\section{Fabrication} To define the structure, we used the ultra high resolution Electron Beam Lithography (EBL). In order to minimize the size of $a$:InO contacts directly adjoin to the structure, we had to implement the EBL process twice with an overlay precision of less than 20 nm between phases: in the first step we produced the inner Ti/Au contacts, followed by the fabrication of the $a$:InO ring (using a second EBL step). Each time thermally oxidized silicon wafer (Si/SiO$_2$ with typical value of the surface roughness less than 1 nm; oxide layer 300 nm and resistivity less than 5 m$\Omega \cdot$cm), was spin-coated with bilayer of poly(methyl methacrylate) (PMMA) electron-beam resist of two different molecular weights. The desired structure was exposed in the resist using a EBL-system JEOL JBX-9300FS. Photolithography was used to prepare four or six-point Ti/Au electrical (outer) contacts. $a$:InO film was $e$-gun evaporated in ultra high vacuum system ($2.5\times10^{-7}$ Torr; Thermionics) from high purity (99.999 \%) In$_{2}$O$_{3}$ pellets in residual O$_{2}$ pressure \nolinebreak $\sim1.5\times10^{-5}$\nolinebreak Torr.  

For structural determination we deposit one more test-sample along with the experimental one. From the scanning electron microscopy (SEM) we conclude that the internal diameter is 50 nm and the external diameter ($d_e$) is 150 nm. The external diameter of the disk (unpatterned film, which we used as a reference) was 320 nm. Accuracy of these measurements was $\pm2$ nm. The SEM images of the obtained structures are shown in the Fig. \ref{fig:semDISK} and \ref{fig:semSTR} (one of four experimental samples exhibiting oscillations is shown). Atomic force microscopy (AFM) images showed the thickness variation about $10\%$, i.e. the thickness is $30\pm3$ nm. $a$:InO is known to form relatively uniform films\cite{hebard1982structural}, so we expect our structures to be uniform as well.

After gentle lift-off, the sample was mounted on the sample holder, electrically connected with Au wire. Finally, the sample was immersed into a Kelvinox TLM (Oxford Instruments Inc.).

We implemented two- and four-probe technique. In the four-probe measurements resistance of the structure included resistance of a small Ti/Au contacts overlapped by $a$:InO contacts. Such pair of contacts was less than 1 $\mu$m$^2$ from each side of the structure, and was caused by the design limitations. The signal from the sample was amplified by low-noise, home-built differential voltage pre-amplifier 
and measured using EG\&G 7265 Lock-in Amplifiers at a frequency of 1.8 Hz. In order to minimize heating of the 
structure, we used low excitation current of 1 nA.

\section{Experiment} 
\begin{figure}[tbh]
\centering
\begin{subfigure}[b]{.22\textwidth}
\includegraphics[width=\textwidth, angle = 0]{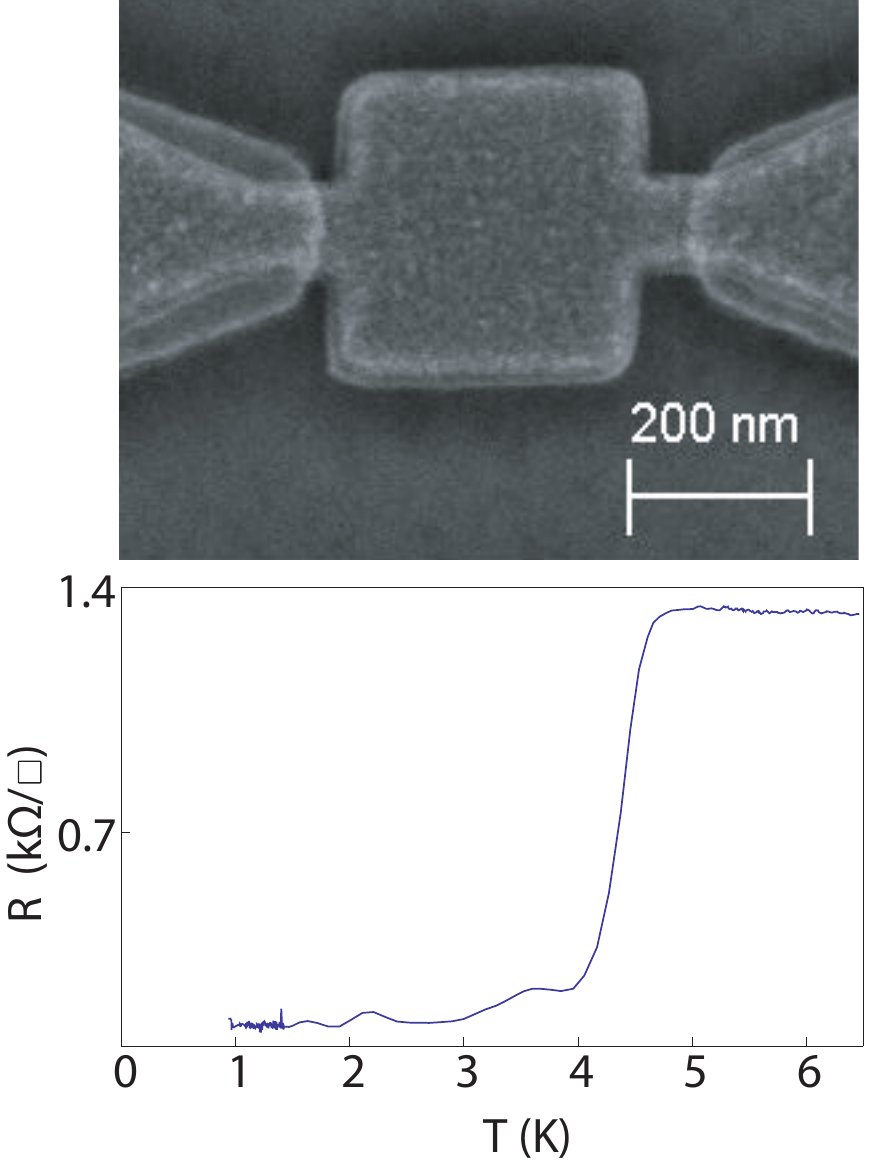}
\caption{}
\label{fig:semDISK}
\end{subfigure}
~ 
\begin{subfigure}[b]{.22\textwidth}
\includegraphics[width=\textwidth]{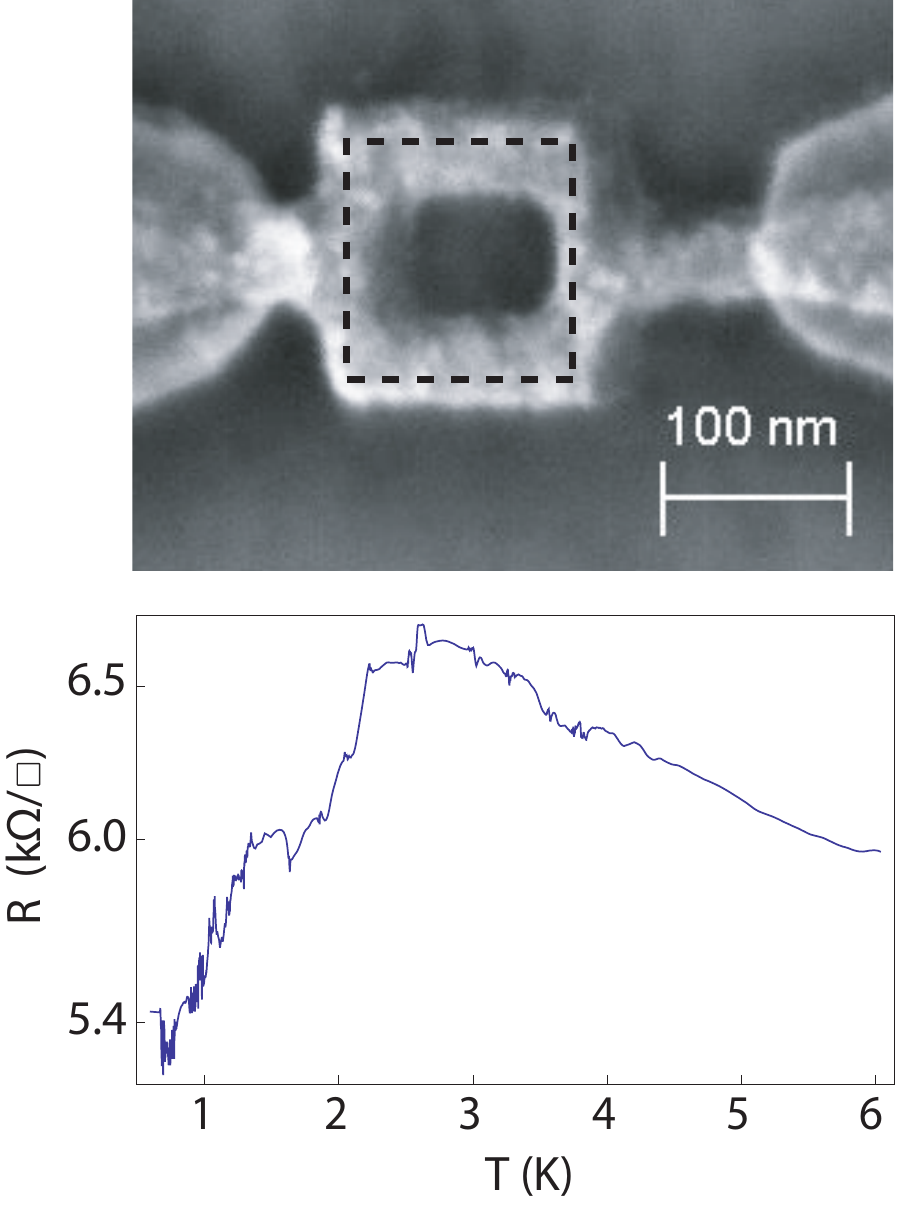}
\caption{}
\label{fig:semSTR}
\end{subfigure}
~ 
\begin{subfigure}[b]{.31\textwidth}
\includegraphics[width=\textwidth]{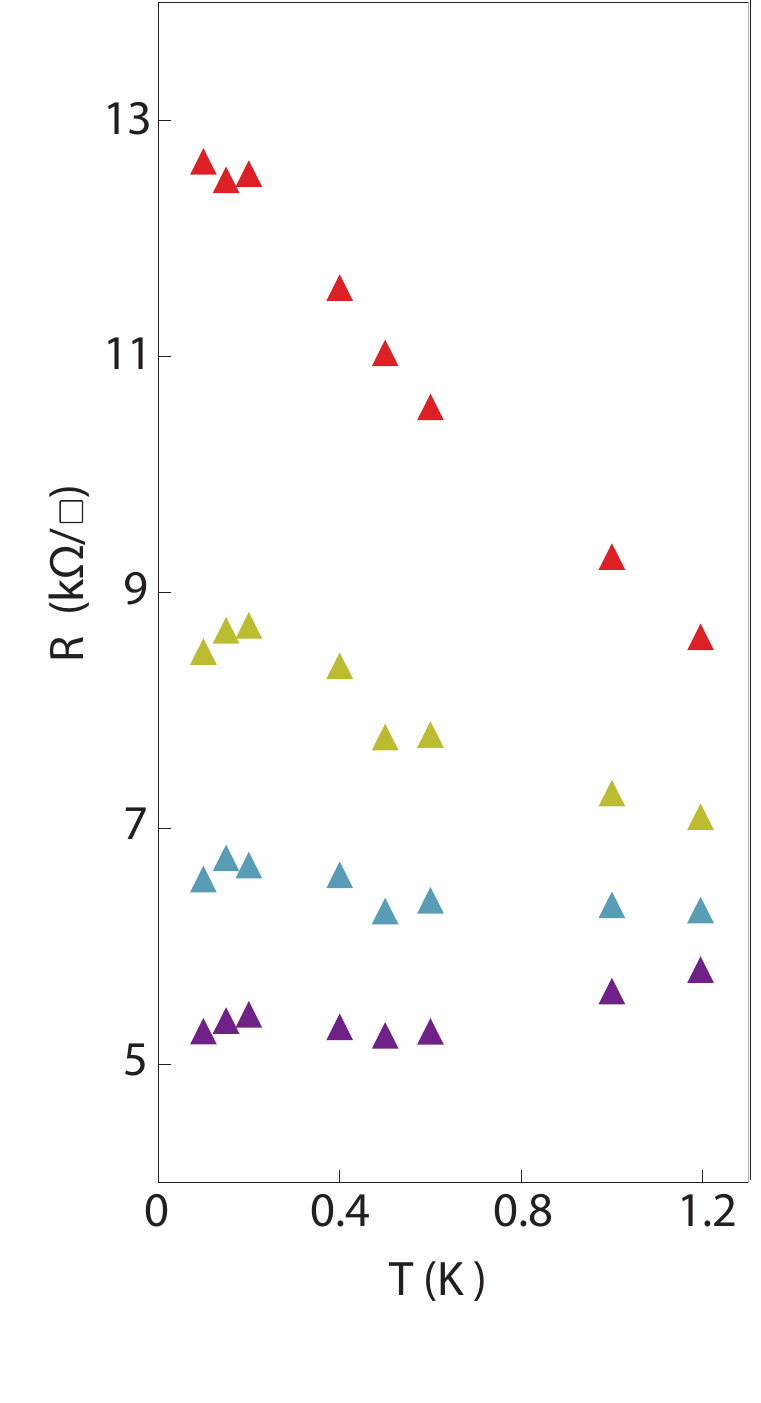}
\caption{}
\label{fig:rtbSTR}
\end{subfigure}
\caption{(color online) SEM-image and $R_{\square}$ vs $T$ at $B=0$ T for the disk (a) and the ring (b). The number of squares for such a ring (5 squares) has been estimated from the geometry of the ring. Dashed line in (b) is the trajectory of a particle of a charge $2e$. (c) $R$ vs $T$ for the ring at $B = 0, 0.7, 1.5, 4.0$ T (from bottom to the top) showing the shift to insulating behaviour.}\label{loading}
\end{figure}

We first measured the dependence of the resistance ($R$) of the $a$:InO disk on temperature ($T$) at $B=0$ T. 
The result is shown in Fig. \ref{fig:semDISK}. As $T$ is lowered below 4 K the resistance drops abruptly 
from 1.4 k$\Omega$ to 50 $\Omega$. In Fig. \ref{fig:semSTR} we present $R$ vs $T$ at $B=0$ T for the ring. Unlike the disk, it does not show an abrupt change in $R$, but a drop of 20\% at $\sim$ 3 K, is most likely due to the (not fully developed) superconducting transition.
We note that despite the sharp drop $R$, it saturates at measurable value and remains finite down to the $T=50$ mK. 
In this regime, the sample demonstrates quadratic positive magnetoresistance at low field turning into a negative magnetoresistance at $B>2$ T. 

Next, we measured $R$ of the ring as a function of $T$. Contrary to the disk and films, it does not show any sudden change in 
the resistance down to the lowest temperature. However, it demonstrates  non-monotonous magnetoresistance, similar to that of the disk. 
We show $R$ vs $T$ traces at different values of magnetic field in Fig. \nolinebreak \ref{fig:rtbSTR}.

\begin{figure}[thb]
\includegraphics[width=0.45\textwidth]{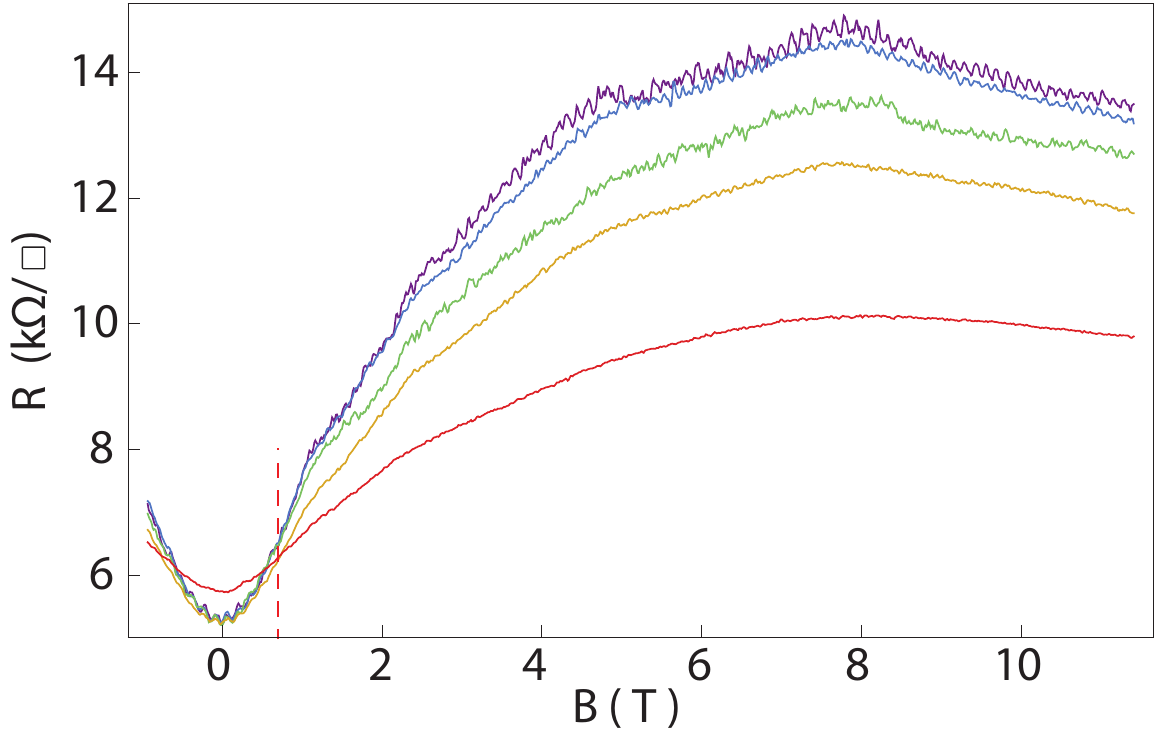}
\caption{(color online) $R$ vs $B$ for temperatures $T= 0.1, 0.2, 0.4, 0.6, 1.2$ K (from top to bottom at $B = 8$ T), 
crossing point $B_c\sim0.8$ T is shown by red dashed line.}
\label{fig:rb}
\end{figure}

On a large scale of $B$, the disk and the ring demonstrate similar behaviour, albeit, in comparison with the disk, the  $R$ vs $T$ dependence at $B=0$ T of the ring is much weaker. They exhibit the high-$B$ phenomenology that we are accustomed 
to in our previous studies of $a$:InO films (see Ref. \citenst{sambandamurthy2004superconductivity}), although, in this case, it is less developed. In Fig. \ref{fig:rb}, we plot $R$ isotherms over our entire $B$ range. 
The crossing point of the isotherms at $B_c= 0.8$ T identifies the 'critical' $B$ of the magnetic field tuned SIT, followed by the prominent magnetoresistance peak at $B = 8$ T. In this experiment we were not able to determine the crossing point in Fig. \ref{fig:rb} better than specifying that it is in the range [0.8 T, 0.9 T]. We believe that relative smallness (compared to measurements on macroscopic films) of the resistance variation with $B$ and $T$ is due to mesoscopic nature of our sample. 
Another effect of the finite size, related to the loop geometry, is clearly seen on the Fig. \ref{fig:rb}: small, about $\sim 1\%$ by 
magnitude, oscillations of resistance as function of magnetic field appear, which will be in the focus of the remainder of this Letter.

\begin{figure}[tbh]
\includegraphics[width=0.45\textwidth]{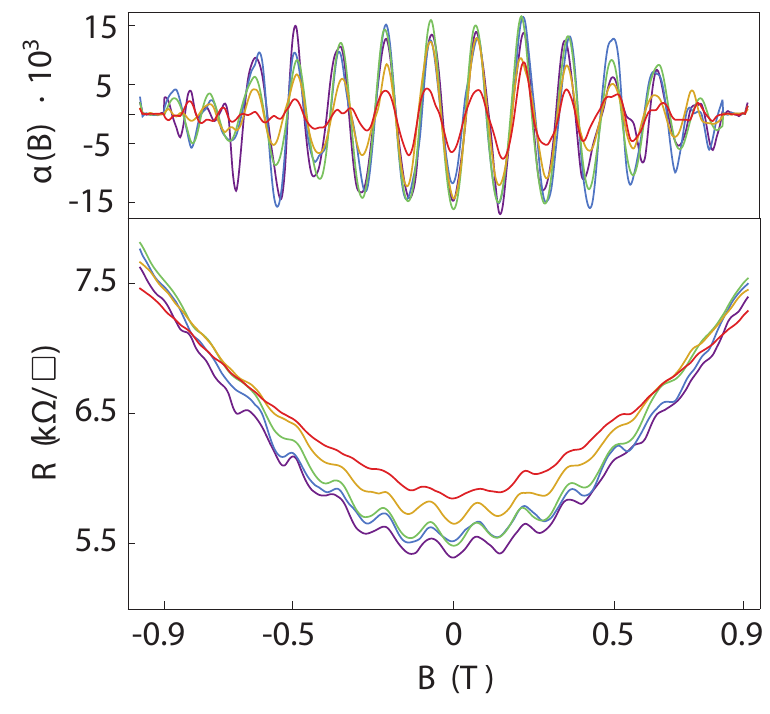}
\caption{(color online) $R$ vs $B$ for the temperatures $T = 0.15, 0.2, 0.5, 0.8$ K (from bottom to the top at $B=0$ T); 
inset: oscillating part $\alpha(B)$ (See text).}
\label{fig:lowB}
\end{figure}

We start our analysis of these oscillations with the region of low $B$. On the plot of $R$ vs $B$ (Fig. \ref{fig:lowB}), more than 
ten oscillations are seen, superimposed on a parabolically rising background. The oscillations period $\Delta B \approx 0.15\pm0.02$ \nolinebreak T 
can be easily read from this figure.  It is independent of $T$, indicating that it is determined by the geometry of the ring. 
Trajectory of a particle of a charge $2e$ encompassing the superconducting flux quantum $\Phi_0=h/2e$ in a field of $0.15$ T is shown on the Fig. \ref{fig:semSTR} and is consistent with 
the flux periodicity in integer units of $\Phi_0$. For better characterization of oscillations, it is convenient to define normalized 
oscillating part $\alpha(B) = (R (B)- R_{s}(B))/R_s(B)$, where $R_s(B)$ is smooth 
part of the $R(B)$ dependence (averaged over several oscillations).

\begin{figure*}
\includegraphics[width=1\textwidth]{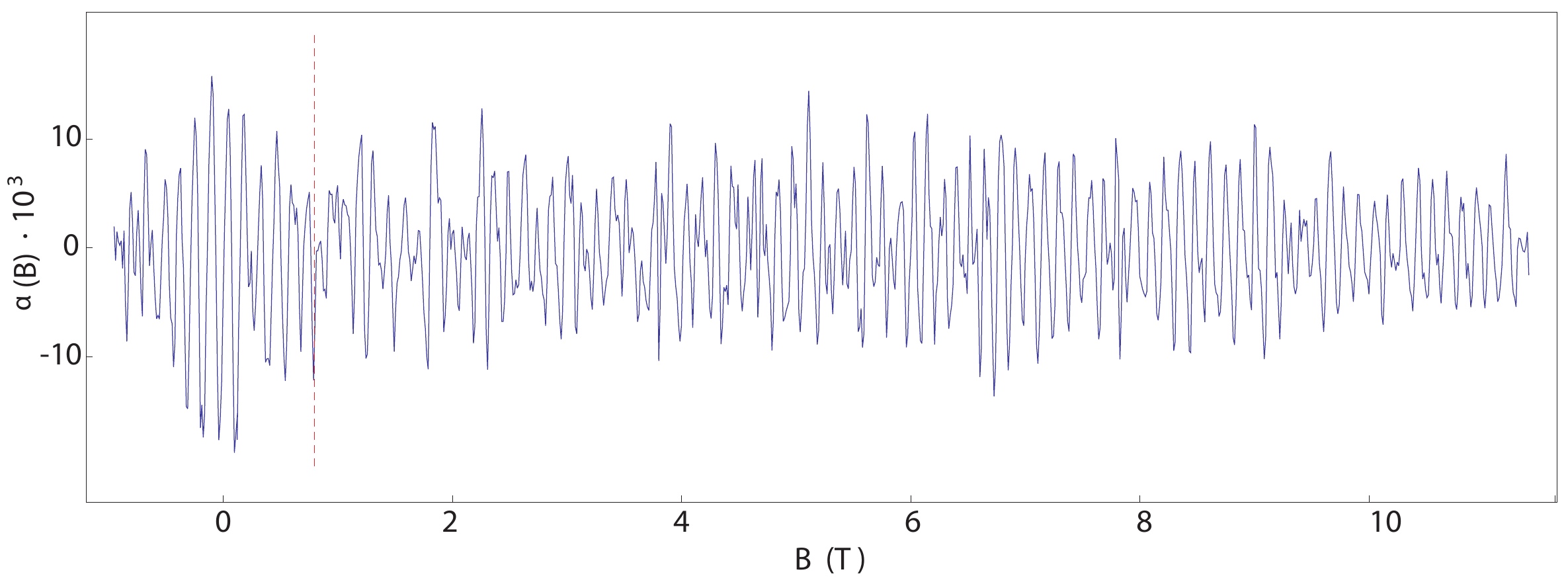}
\caption{(color online) oscillating part $\alpha(B)$ for $T=0.15$ K, red dashed line shows $B_c\sim0.8$ T}
\label{osc}
\end{figure*}

\begin{figure}[tbh]
\includegraphics[width=0.45\textwidth]{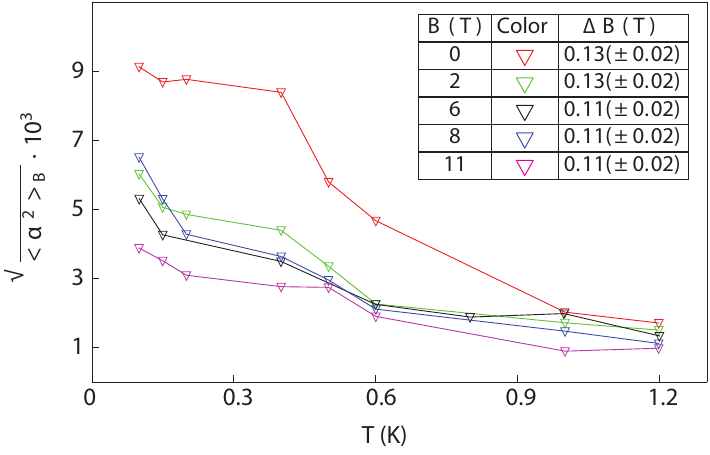}
\caption{(color online) Amplitude of the oscillations vs temperature for different $B = 0, 8, 2, 6, 11 $ T.
Inset: the period of the oscillations as determined by counting the peaks in the interval of 1 T in the several ranges of $B$.}
\label{TDep}
\end{figure}

Our central result is related to the behaviour of $\alpha(B)$ at high $B$. It is presented in Fig. \ref{osc}, 
where we plot $\alpha(B)$ of our ring for the entire range of $B$ at $T=150$ mK. Oscillations are clearly visible throughout the range, 
up to our highest $B$. This result is quantified in a table, shown inset on the Fig \ref{TDep}.,
where, we show the period of the oscillations as determined by counting the peaks in the interval of 1 T on several ranges of $B$.
Different rows correspond to different ranges of $B$: the first row, for example, is for the range [-0.5 T, 0.5 T]. 
It is clear that the oscillations have similar periodicity at different values of $B$. 

Finally, we characterize the $T$-dependence of the amplitude of the oscillations. 
As a quantity characterizing the amplitude of the oscillations, we choose $\sqrt{\left<\alpha^2\right>_B}$, where averaging over
the entire range of $B$ is implied (the results are qualitatively the same if averaging over other ranges are performed). The $T$-dependence of this quantity is shown in Fig. \ref{TDep} . It is consistent with our intuition: with increasing $T$, coherence length of the Cooper pairs decreases and oscillations disappear at $T\sim 1.2$ K.

\begin{figure}[tbh]
\centering

\begin{subfigure}[b]{.35\textwidth}
\includegraphics[width=\textwidth]{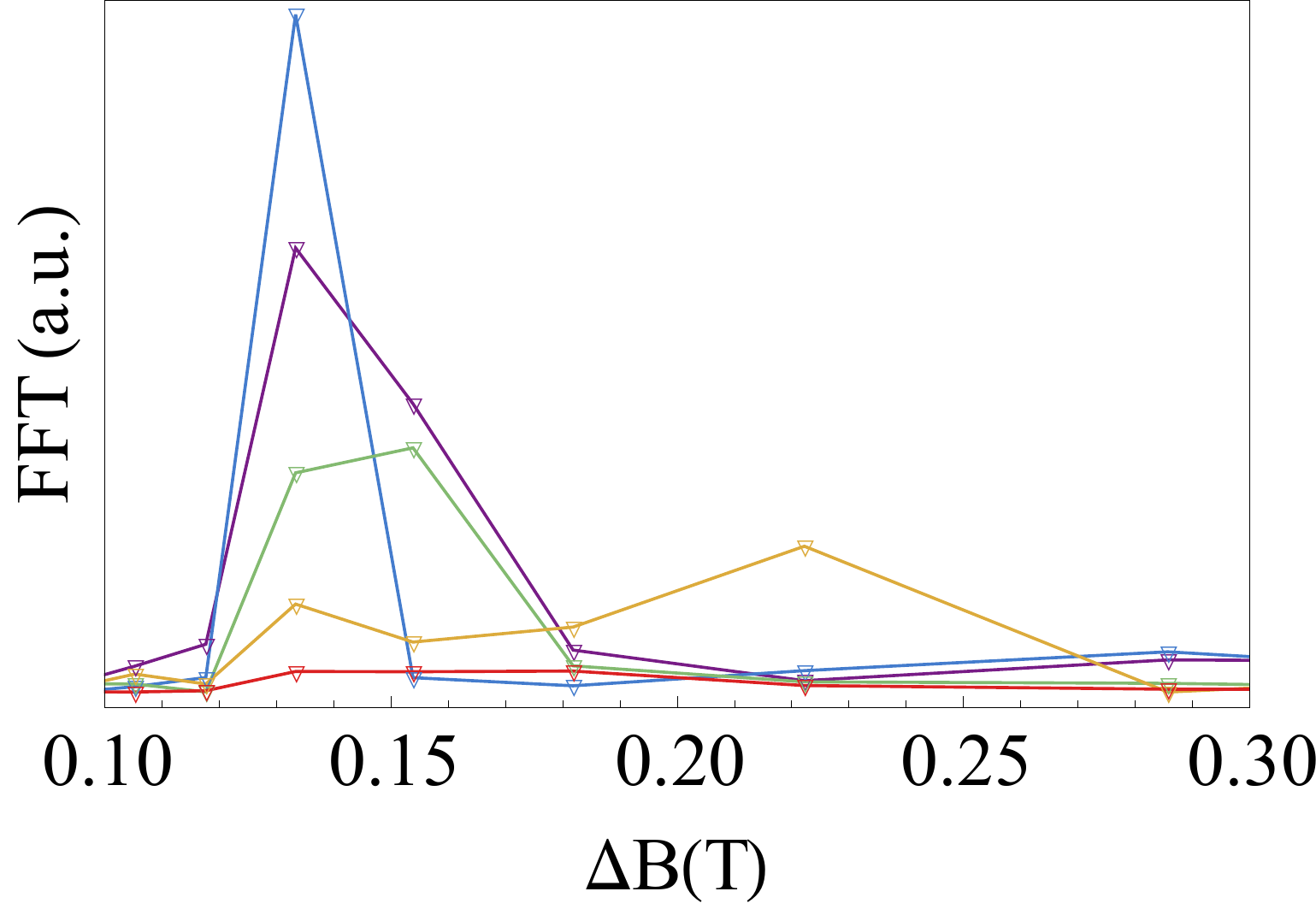}
\caption{} \label{fig:FFTRvsT}
\end{subfigure}
~ 
\begin{subfigure}[b]{.35\textwidth}
\includegraphics[width=\textwidth]{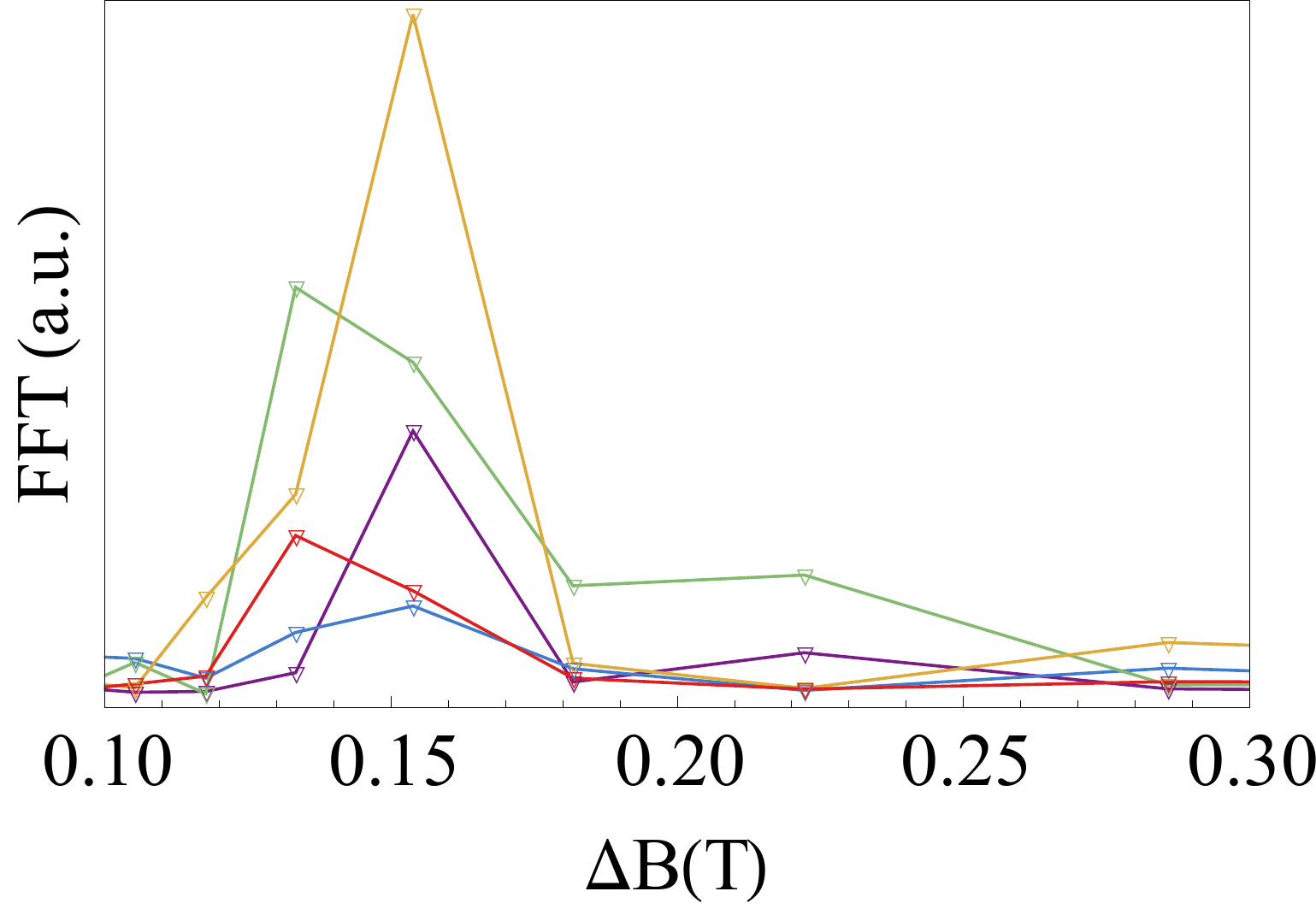}
\caption{} \label{fig:FFTRvsB}
\end{subfigure}

\caption{
(color online) (a) FFT for temperatures $T = 0.15, 0.1, 0.2, 0.5$ K (from top to the bottom at $\Delta B \sim 0.15$ T). (b) FFT for $B=8, 6, 0, 10, 2$ T, from top to the bottom at $\Delta B \sim 0.13$ T and $T=0.1$ K. The field window is 1 T: i. e., purple line is for the range [-0.5, +0.5] T.} \label{fig:FFT}
\end{figure}

In order to further quantify this result, we performed Fourier Transform (FFT) of the oscillating contribution. The result is demonstrated in Fig. \ref{fig:FFT}. The main conclusion of this analysis is that the period of the oscillations is the same at different $T$ and $B$. In Fig. \ref{fig:FFTRvsT} we show the spectra as function of $T$. It is clear that the period of oscillations remains $\Phi_0$ for different temperatures. At the same time, the amplitude of the dominant peak is strongly $T$ dependent. In order to quantify how the oscillatory properties change with increasing $B$, we perform a series of FFTs at different sub-regions of field (see Fig. \ref{fig:FFTRvsB}), where several curves correspond to FFT of signals in different ranges of $B$: the lowest curve, for instance, shows the data in the range [-0.5, 0.5] T. This plot demonstrates that oscillations have similar periodicity in different ranges of the magnetic field (within \nolinebreak an error $\sim 0.02$ T).

\section{Discussion}
Our main observation are magnetoresistance oscillations of a constant period throughout the available interval of $T$ and $B$, consistent with a flux periodicity, corresponding to elementary charge $2e$. We argue below that this magnetoresistance is most likely due to electron-electron interaction in the Cooper channel,  that is, undeveloped Cooper pairing.

We are aware of two physically distinct mechanisms that can lead to such oscillatory magnetoresistance:
the electron-electron interaction in the Cooper channel and weak (anti)localization (WL). The first effect is expected to be most prominent close to the 
superconducting transition, as is measured in the Little-Parks scheme\cite{little1962observation}. In the vicinity of $T_c$ the resistance of the ring 
is determined by thermal phase slips, which are influenced by magnetic flux penetrating the ring. Detailed theoretical analysis of this effect 
was performed in Ref. \citenst{sochnikov10}, where it was demonstrated that  periodic flux-dependence 
of activation energy of the phase slips in such a ring allows to explain the magnitude of experimentally observed oscillations in the vicinity of superconducting transition of the LaSrCuO rings \cite{sochnikov2010large}. For the metallic regime outside the transition 
region in the vicinity of $T_c$ Kulik \cite{kulik1972} predicted, based on the Ginzburg-Landau approach, that this effect should also be 
noticeable. It appears due to the presence of fluctuating Cooper pairs, which can be rather long-lived, $\tau_{GL}=\frac{\pi}{8T\ln T/T_c}$ and in this respect
is due to paraconductivity\cite{aslamazov1968effect}. Later, Larkin demonstrated \cite{larkin1980reluctance} that further away from the transition 
Maki-Thompson correction would be dominant in magnetoresistance and, hence, will give dominant contribution in Little-Parks type of \nolinebreak oscillations. 

Interestingly, in the experiment of Shablo\cite{shablo1974quantization}, where oscillations of the resistance in the normal state were first observed,
they were attributed to paraconductivity, and only later it became clear that it is more realistic that they are actually related to weak localization, which was not 
well understood. It gives another possible contribution to the observed oscillations\cite{alt81}, see also experimental study in Ref. \citenst{sharvin81}. 
This effect is not associated with electron-electron interaction, but its phenomenology is similar to that of interaction-induced one. Interestingly, 
this effect was also predicted to exist in the hopping conductivity regime \cite{nguyen1985tunnel}, but the low field magnetoresistance has a negative sign - opposite to that observed in our measurements.

As was stressed already in the seminal work of Ref. \citenst{alt81}, the amplitude of the oscillations in metals is usually determined by a factor 
$\gamma-\beta(T)$, where $\gamma$ is coming from WL part and depends on the symmetry class of the system ($\gamma=1$ for weak spin-orbit impurity scattering, $\gamma=-1/2$ for opposite case)
and $\beta(T)$ is an effective constant of Cooper interaction\cite{larkin1980reluctance}. These two effects, although having the same periodicity, 
have rather different spatial scales: single-particle coherence length $L_{\phi}$ 
and coherence length of the Cooper pair $L_{\xi}$. Since we do not have any reliable estimate for $L_{\xi}$ in our sample, 
we will concentrate on ruling out the possibility of WL origin of the effect.
First we consider the ability of single electrons to maintain coherence on the size of the sample. At the temperatures of our experiment, the main source of dephasing is expected to be electron-electron interaction. 
According to Ref. \citenst{AAK82}, coherence length $L_\phi = \sqrt{D \tau_{\varphi}}$ can be estimated 
from dephasing rate $\frac{1}{\tau_{\varphi}} =\frac{T}{2\pi g}\ln(2\pi g)$. Estimating the diffusion coefficient as 
$D \approx 1$ cm$^2$/s and $g=\hbar\sigma/e^2 \approx 0.7$ we find for $T=0.2$ K the values of $\tau_{\varphi} \approx 100$ ps 
and $L_{\phi} \approx 90$ nm, which is much smaller than the circumference of our ring. Additionally, the magnetic field not only imposes a phase on the interfering electrons, but also induces mass into the Cooperon. This effect becomes more pronounced with the growth of the width w of the ring. As shown in Ref. \citenst{al1982observation}, the effective dephasing length for the WL-induced oscillations $\bar{L}_{\phi}$ is determined by $1/\bar{L}_{\phi}^{2} = 1/L_{\phi}^{2} + \frac{1}{3}(\frac{weH}{\hbar c})^2$. This effect imposes additional restriction on the number of oscillations, which can be seen in the experiment as follows: $N_{osc} \sim d_e /w$, in our case $N_{osc} \sim 3$, while we resolve over 90 oscillations.

\section{CONCLUSIONS}

These arguments allow to exclude WL for the description of the observed effect. We emphasize that, while we defer an attempt for a theoretical explanation of our observation to a future publication, we nevertheless stress that in all likelihood it is rooted in the stability of Cooper pairing deep in the insulating regime. This is especially interesting because our ring does not show strong superconducting trend. More detailed study of these oscillations can shed more light on the role of the Cooper interaction in mesoscopic $a$:InO rings.

\section*{ACKNOWLEDGMENTS} We acknowledge useful dissociations with M. V. Feigel'man, A. Finkel'stein, Y. Oreg, B. Spivak and C. Marrache-Kikuchi, S. Levit, D. Melnikov, O. Raslin, V. Hanin, P. Khatua, A. V. Sevast'yanov and R. Yuval.
This work was supported by the Minerva foundation with funding from the Federal German Ministry for Education and Research. 
KT is supported by the Paul and Tina Gardner fund for Weizmann-TAMU collaboration.

\bibliography{rings}

\end{document}